\documentclass[%
superscriptaddress,
preprint,
 amsmath,amssymb,
 aps,
]{revtex4-2}

\usepackage{graphicx}
\usepackage{dcolumn}
\usepackage{bm}
\usepackage{multirow}
\usepackage[version=3]{mhchem}

\usepackage{amsmath,esint}
\usepackage{verbatim}
\usepackage{spverbatim}
\RequirePackage[normalem]{ulem}
\RequirePackage{color}\definecolor{RED}{rgb}{1,0,0}\definecolor{BLUE}{rgb}{0,0,1}\definecolor{GREEN}{rgb}{0,1,0}

\usepackage{hyperref}
\usepackage{fancyhdr}
\begin{document}


\title{WloopPHI: A tool for \textit{ab initio} characterization of Weyl semimetals}

\author{Himanshu Saini}
\author{Magdalena Laurien}
\affiliation{Department of Materials Science and Engineering, McMaster University, 1280 Main Street West, Hamilton, Ontario L8S 4L8, Canada}

\author{Peter Blaha}
\email[P.B.: ]{ORCID: 0000-0001-5849-5788}
\affiliation{Institute of Materials Chemistry, Vienna University of Technology, Getreidemarkt 9/165-TC, A-1060 Vienna, Austria}

\author{Oleg Rubel}%
\email[O.R. email: ]{rubelo@mcmaster.ca, ORCID: 0000-0001-5104-5602}
\affiliation{Department of Materials Science and Engineering, McMaster University, 1280 Main Street West, Hamilton, Ontario L8S 4L8, Canada}

\date{\today}

\begin{abstract}
\texttt{WloopPHI} is a Python code that expands the features of \texttt{WIEN2k}, a full-potential all-electron density functional theory package, by the characterization of Weyl semimetals. It enables the calculation of the chirality (or  ``monopole charge") associated with Weyl nodes and nodal lines. The theoretical methodology for the calculation of the chirality is based on an extended Wilson loop method and a Berry phase approach. We validate the code using \ce{TaAs}, which is a well-characterized Weyl semimetal, both theoretically and experimentally. Afterwards, we applied the method to the characterization of \ce{YRh6Ge4} and found two sets of Weyl points (ca. 0.2~eV below the Fermi energy) together with a topological nodal line (protected by mirror symmetry) crossing the Fermi energy and mapped their chiralities.
\end{abstract}

\maketitle


\section{Introduction}

Advancements in the field of topological materials \cite{RevModPhys.82.3045,RevModPhys.83.1057,Bansil_RoMP_88_2016,Basov_NM_16_2017,Armitage_RMP_90_2018} have led to the prediction and discovery of new phenomena, such as ultrahigh magnetoresistance \cite{Liang2015}, chiral anomaly \cite{PhysRevX.5.031023,Zhang2016,Xiong413,NIELSEN1983389,PhysRevB.88.104412}, the intrinsic anomalous Hall effect \cite{Nayake1501870,Suzuki2016,Wang2018,Liu2018,Nakatsuji2015}, and exotic Fermi arc surface states \cite{PhysRevB.83.205101,Xu613}. The spectrum of topological materials has expanded from insulators to metals/semimetals \cite{Zhang2016,Weng_2016,weng2016topological,Dzsaber_PNAS_118_2021}. Weyl semimetals (WSMs) are a topological variant of gapless semimetals. WSMs are characterized by the presence of topological Fermi arc surface states that connect Weyl nodes of opposite chirality \cite{Young_PRL_108_2012,Xu613,PhysRevX.5.031013,PhysRevB.83.205101,PhysRevLett.107.186806,PhysRevX.5.011029,Lv2017}. Weyl points occur in materials as a result of breaking either time-reversal or spacial inversion symmetry (but not both) for a Dirac nodal ring in the presence of spin-orbit coupling (SOC)  \cite{Young_PRL_108_2012,Liu_PRB_90_2014,PhysRevX.5.011029}.

\begin{figure}
	\includegraphics{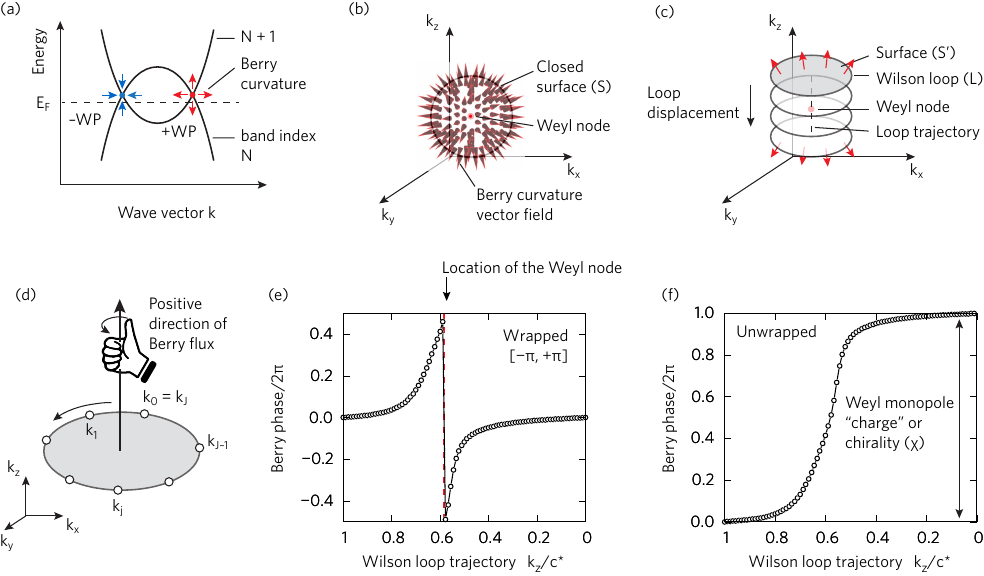}
	\caption{(a)~Weyl nodes of opposite chirality at the intercept of two bands. Arrows indicate the source (red) and sink (blue) of the Berry curvature $\nabla_{\bm{k}} \times \bm{A}_n$. (b)~The node can be viewed as a monopole with a chiral charge (positive node is shown with the chirality $\chi=+1$). The charge is determined by a Berry flux through a closed surface $S$ that encloses the Weyl node. (c)~Construction of a series of Wilson loops that enclose one Weyl node. There is a Berry phase $\phi$ associated with each loop that represents a Berry flux through the surface $S'$ of the loop. (d) Use of a right-hand rule in definition of a positive direction of the Berry flux through the loop. (e) Evolution of Berry phase (wrapped in the range $[-\pi,+\pi]$) as a function of the loop's position along the $k_z$ direction.  The discontinuity (marked with an arrow) corresponds to a Weyl point. (f) A finite Berry phase (unwrapped) accumulates as the Wilson loop traverses through the BZ. The accumulated phase represents the chirality $\chi=+1$ of the Weyl point.}\label{fgr:Figure-1}
\end{figure}

The chirality of Weyl nodes can be characterized as the Berry flux through a closed surface $S$ in reciprocal $k$ space (Fig.~\ref{fgr:Figure-1}a,b) that surrounds the Weyl node in question \cite{PhysRevX.5.011029}
\begin{equation}\label{eq:Berry-flux-gamma_n}
	\gamma_n = \oiint_S (\nabla_{\bm{k}} \times \bm{A}_n) \cdot d\bm{S}.
\end{equation}
Here $n$ is the band index for the lower of the two bands at the crossing (i.e., $n=N$ in Fig.~\ref{fgr:Figure-1}a) and $\bm{A}_n$ is a Berry connection \cite{Berry_PRSLSA_392_1984}
\begin{equation}\label{eq:Berry-connection-A_n}
	\bm{A}_n(\bm{k}) = i \langle u_{n\bm{k}} | \nabla_{\bm{k}} | u_{n\bm{k}}  \rangle,
\end{equation}
with $u_{n\bm{k}}$ being the cell-periodic part of the Bloch wave function $\psi_{n\bm{k}}(\bm{r})=u_{n\bm{k}}(\bm{r}) e^{i\bm{k}\cdot \bm{r}}$, and the term $\nabla_{\bm{k}} \times \bm{A}_n$ representing the Berry curvature vector field. The topological properties of a Weyl point are captured by its chirality
\begin{equation}\label{eq:chirality-number-(general)-chi_n}
	\chi = \gamma_n/2\pi.
\end{equation}
Most commonly Weyl points come in pairs of opposite chirality or opposite ``monopole charge" $\pm 1$ as stated by the no-go theorem \cite{NIELSEN1981173,NIELSEN198120}. However, in general, bands crossings with a greater monopole charge magnitude (such as 2, 3, etc.) are also possible \cite{fang_multi-weyl_2012}. Here we refrain from using a term ``Chern number" as a synonym for chirality of individual Weyl points, even though such a trend exists in the literature \cite{Soluyanov_N_527_2015,Hasan_ARCMP_8_2017,yan2017topological}.   The intent is to avoid a possible confusion with another Chern number defined as an integral of the Berry curvature over the whole Brillouin zone and occupied bands that governs an anomalous Hall conductivity (Ref.~\citenum{Vanderbilt_Berry_2018}, chap.~5).

\texttt{WannierTools} \cite{Wu_CPC_224_2018} and \texttt{Z2Pack} \cite{Gresch_PRB_95_2017,Soluyanov_PRB_83_2011} are open-source packages that offer comprehensive collections of tools for the theoretical characterization of topological materials, including WSMs. In both packages, the Weyl point's chirality is evaluated by constructing a small sphere enclosing the node in question. To determine enclosed Weyl point's chirality they track an evolution (winding) of the sum of hybrid Wannier charge centers computed on loops (contracted on the sphere by analogy with parallels of the globe) as a function of the polar angle. However, there is one important difference between these two packages.  \texttt{WannierTools} provides an \textit{indirect} characterization of \textit{ab initio} electronic structure mediated by a tight-binding framework  generated by \texttt{Wannier90} \cite{MOSTOFI2008685}. \texttt{Z2Pack}, on the other hand, works at all levels of materials modeling including a \textit{direct} interface with \textit{ab initio} calculations \cite{Gresch_PRB_95_2017}.  \texttt{Z2Pack} also requires \texttt{Wannier90}, but only to generate a list of nearest-neighbour $k$ points (a \texttt{seedname.nnkp} file), which is later used by a first-principles
code to calculate overlap projections and eventually the Wannier charge centers.

Here we introduce \texttt{WloopPHI}, an open-source code for the characterization of Weyl semimetals. Unlike  \texttt{WannierTools}, \texttt{WloopPHI} can be run directly from first principles without constructing a Wannier Hamiltonian or using \texttt{Wannier90}. \texttt{Z2Pack} can also be run from first principles, i.e., without Wannierization, however it required \texttt{Wannier90} to complete the calculations for Wannier charge centers. Furthermore, \texttt{Z2Pack} has an interface with several plane wave codes, but not with full potential all-electron codes such as \texttt{WIEN2k}.

For \texttt{WloopPHI}, we propose an alternative methodology to characterize Weyl points and calculate their chirality. Instead of constructing a closed surface $S$, we select a small open surface $S'$ defined by a closed loop  $L$ in reciprocal space (Fig.~\ref{fgr:Figure-1}c). The Berry flux through this surface can be found using Stokes' theorem
\begin{equation}\label{eq:Berry-flux-Stokes}
	\phi_n = \iint_{S'} (\nabla_{\bm{k}} \times \bm{A}_n) \cdot d\bm{S}' = \oint_L \bm{A}_n \cdot d\bm{k}.
\end{equation}
The main advantage of this approach is that we do not need to know the Berry curvature vector field. Further, our approach does not require any gauge-fixing condition. Integration of the Berry connection on a closed loop of $k$ points in reciprocal space (Fig.~\ref{fgr:Figure-1}d) is well established and can be performed directly in \textit{ab initio} electron structure codes as outlined by \citet{King-Smith_PRB_47_1993}. We are interested in how the Berry flux through $S'$ (same as the Berry phase on the loop) changes as the loop moves along a trajectory that encloses a Weyl node (Fig.~\ref{fgr:Figure-1}c). The loop trajectory can be chosen such that it evolves in a ``negative" direction defined by a right-hand rule (Fig.~\ref{fgr:Figure-1}d). Since Weyl points represent either a source or a sink of the Berry curvature vector field, its flux through the loop $S'$ will be small when the loop is far away from the Weyl node and will gradually increase as the loop approaches the node (Fig.~\ref{fgr:Figure-1}e). The presence of a discontinuity in the Berry phase will signify the location of a Weyl node. The chirality is expressed as the winding number of the Berry phase along the trajectory of the loop (Fig.~\ref{fgr:Figure-1}f) of the enclosed Weyl node(s) \cite{Liu_PRB_90_2014}. The advantage of enclosing a Weyl node in the cylindrical shape (vs a spherical surface used in \texttt{Z2Pack} or \texttt{WannierTools}) is that the exact position of the Weyl node along the height of the cylinder does not have to be known in advance; the downside is a higher computational intensity of calculations since more loops are to be constructed (in general) to map the height.

We refer to our method as an extension of the ``Wilson loop" approach to acknowledge its connection with the calculation of a gauge field on a closed path suggested by \citet{Wilson_PRD_10_1974}. However, it should not be confused with another variant of the Wilson loop technique \cite{PhysRevB.84.075119,Taherinejad_PRB_89_2014} (also known as an evolution of Wannier charge centres) that is used for calculation of the  $\mathbb{Z}_2$ topological invariant. The latter implies calculation of the Berry phase on a one-dimensional (1D) string  of $k$ points along a specific direction in the Brillouin zone (BZ) (analogous to calculation of electronic polarization in solids \cite{King-Smith_PRB_47_1993}). The closed loop character of this one-dimensional string is ensured by a periodicity of the BZ. The evolution of Wannier charge centres is applied to WSMs to reveal their $\mathbb{Z}_2$ topological invariant \cite{Sheng_PRB_90_2014,Sun_PRB_92_2015,PhysRevX.5.011029}, albeit without giving any details on the chirality of individual Weyl points. The Wilson-loop and hybrid-Wannier-centres method has also been modified for type-II WSMs by \citet{Soluyanov_N_527_2015} in such a way that circular Wilson loops are constructed on a small sphere around a Weyl point to obtain its chirality.

\texttt{WIEN2k} \cite{Blaha_WIEN2k_2018,Blaha_JCP_152_2020}, an all-electron  density functional theory \cite{Hohenberg_PR_136_1964,Kohn_PR_140_1965} package, is selected as platform for the implementation of our method. We use the \texttt{BerryPI} code \cite{AHMED2013647} in conjunction with \texttt{Wien2wannier} \cite{Kunes_CPC_181_2010} (both implemented in the \texttt{WIEN2k} package) for the calculation of the Berry phase on the Wilson loop. To verify our methodology, we characterize the well-known topological material TaAs \cite{Xu613,PhysRevX.5.031013,PhysRevX.5.011029} and subsequently analyse the Weyl points of a recently discovered material, \ce{YRh6Ge4} \cite{doi:10.1002/zaac.201300369,PhysRevB.101.035133,PhysRevB.98.045134}.

\section{Method}
\subsection{Chirality of a Weyl point}

We used an expanded Wilson loop method to determine the chirality of Weyl points. A Wilson loop is defined as an arbitrary closed $k$-point path in the BZ (Fig.~\ref{fgr:Figure-1}d). Wave functions evaluated around this path acquire a total Berry phase $\phi(k)$. Here $k$ refers to a coordinate on the loop trajectory rather than a $k$ point on the loop itself. Further, a series of parallel Wilson loops are constructed such that their trajectory encloses a Weyl point (or points) of interest. The chirality corresponds to the winding number of the calculated Berry phase $\phi(k)$ along the trajectory. 

A general expression for the Berry phase on a Wilson loop for a manifold of bands in a range $n_a \ldots n_b$ is determined using \texttt{BerryPI} following the numerical method outlined in Ref.~\citenum{King-Smith_PRB_47_1993}
\begin{equation}\label{eq:Berry-flux-numerical}
	\phi_{n_a : n_b} = \text{Im}
	\left[
		\ln \prod_{j=0}^{J-1} \det \mathbb{M}_{l \times l}(\bm{k}_j,\bm{k}_{j+1})
	\right].
\end{equation}
Here $\mathbb{M}$ is the overlap matrix of the size $l^2=(1+n_b - n_a)^2$, and the product index $j$ runs over all $k$ points on the Wilson loop (Fig.~\ref{fgr:Figure-1}d). Computation of the overlap integral $\mathbb{M}_{mn}(\bm{k}_j,\bm{k}_{j+1})=\langle u_{m\bm{k}_j} | u_{n\bm{k}_{j+1} } \rangle$ between two cell periodic parts of the Bloch function is performed by \texttt{Wien2wannier}. It is sufficient to use only one band in the calculations of the Berry phase ($n_a=n_b=N$ or $n_a=n_b=N+1$, but not both $n_a=N$, $n_b=N+1$). Having a range (e.g., $n_a=1$, $n_b=N$) can be potentially beneficial in case we would like to exclude crossing between bands $N-1$ and $N$ that could accidentally fall within the Wilson loop. In the latter case we can truncate semi-core electrons by selecting an appropriate bottom band range $n_a$ to save computational time.

There are two different sign conventions for the Berry phase (see Sec.~3.2 in Ref.~\citenum{Vanderbilt_Berry_2018}) depending on how the Berry curvature is defined. The convention adapted in Eq.~\eqref{eq:Berry-flux-numerical} implies a standard mathematical definition of the curl with the $z$-axis component of the Berry curvature expressed as $(\nabla_{\bm{k}} \times \bm{A}_n)_z=\partial_y A_x - \partial_x A_y$, which is also compatible with the right-hand rule. Alternatively, the terms $\partial_{\alpha} A_{\beta}$ can be interchanged, \textit{e.g.} as in Ref.~\citenum{Wu_CPC_224_2018}, leading to a reversal of the sign for the Berry curvature and also for the phase in Eq.~\eqref{eq:Berry-flux-numerical}. An implication of the different sign convention is reversal of the sign for the chirality of Weyl nodes. The different conventions do not pose a difficulty for the characterization of WSMs, because predictions of observables (Fermi arcs) rely on the \textit{difference} in the  polarity between Weyl nodes.

Figure~\ref{fgr:Figure-1}c,d show a schematic diagram of the method implemented. The Berry phases $\phi(k)$ computed for individual Wilson loops along the trajectory are wrapped in the range of $[-\pi, +\pi]$ (see Fig.~\ref{fgr:Figure-1}e). If the trajectory contains a Weyl point, a discontinuity in the Berry phase evolution along the trajectory will appear. The discontinuity is explained by an inversion of the Berry flux direction upon the loop plane passing through the Weyl point. The number of discontinuities along the loop trajectory corresponds to the number of Weyl points. To determine the chirality $\chi$ of an individual Weyl point, we unwrap $\phi(k)$ (Fig.~\ref{fgr:Figure-1}f) and evaluate the phase difference between the final and initial positions on the loop trajectory
\begin{equation}\label{eq:chirality-num}
	\chi = \frac{1}{2\pi}\text{unwrap}[\phi({k_\text{fin}}) - \phi(k_\text{init})].
\end{equation}
The definition of the initial and final position implies that the loop evolves along the \textit{negative} direction as defined by the right-hand rule (Fig.~\ref{fgr:Figure-1}d). The unwrapping is performed by the python \texttt{numpy.unwrap} function, which minimizes discontinuities in the array of phases $\phi(k)$ by adding or subtracting a multiple of $2\pi$.

\subsection{Program implementation}

\texttt{WloopPHI} is a python code for \textit{ab initio} calculation of the Berry phase along a series of Wilson loops. It works in a package with \texttt{WIEN2k}, \texttt{Wien2wannier}, and \texttt{BerryPI}. \texttt{WloopPHI} became a part of \texttt{WIEN2k} standard distribution starting with the version 21.1 released on Apr 2021. At the same time, we offer a current development version of \texttt{WloopPHI} via a GitHub repository \cite{BerryPI-GitHub} that captures any work-in-progress and bug fixes prior to the next \texttt{WIEN2k} release.

\texttt{WloopPHI} requires one input file which contains information about the \texttt{WIEN2k} case directory, the number of intermediate Wilson loops on the trajectory, the upper band index $N$, and the list of $k$ points ($\bm{k}_1,\bm{k}_2,\ldots,\bm{k}_{J-1}$, Fig.~\ref{fgr:Figure-1}d) that compose the Wilson loop in the initial and final parts of the trajectory (intermediate loops will be interpolated).

The execution of the code is done by invoking \texttt{WloopPHI} with one argument, which is the Wilson loop input file name (say \texttt{Wloop.in}). This argument is mandatory. Optional arguments include: \texttt{-sp} which allows spin-polarized calculation, \texttt{-orb} which allows for additional orbital potentials (local density approximation + Hubbard U or exact exchange for correlated electrons), and \texttt{-p} for parallel calculation (requires \texttt{.machines} file as described in the WIEN2k users guide \cite{Blaha_WIEN2k_2018})
\begin{verbatim}
	python /path/to/WIEN2k/SRC_BerryPI/BerryPI/WloopPHI.py Wloop.in [-sp] [-orb] [-p]
\end{verbatim}
The spin-orbit coupling is activated implicitly. The code generates one output file, \texttt{PHI.dat}, and associated figures. The output file contains four columns: The Wilson loop trajectory evolution coordinate $k$, the computed ``raw" Berry phases $\phi(k)$, the Berry phases wrapped in the range $[-\pi,+\pi]$, and the unwrapped Berry phases. The last column contains all the information needed for computing the chirality using Eq.~\eqref{eq:chirality-num}. 

To find the coordinates in the first column, we calculate the distance between the first loop and the last loop and divide it by the specified number of Wilson loops which gives the distance between subsequent Wilson loops along the trajectory, $\Delta k$. The $\Delta k$ is used to find the intermediate points along the evolution direction of the Wilson loop which is identical for all k-points in the Wilson loop
\begin{equation}\label{eq:Evolution-Coordinates}
\Delta k = \frac{\sqrt{(k_{x_\text{i}}-k_{x_\text{f}})^2 + (k_{y_\text{i}}-k_{y_\text{f}})^2 + (k_{z_\text{i}}-k_{z_\text{f}})^2}}{n_{\text{wl}}-1},
\end{equation}
where $\bm{k}_\text{i}$ and $\bm{k}_\text{f}$ stand for the initial and final coordinates of the first k-point on the Wilson loop, $n_\text{wl}$ is the number of Wilson loops.

An example output file for \ce{TaAs} for one Weyl point is given below. For this calculation, the Wilson loop was moved along the $k_z$ direction from $k_z = 1$ to $k_z = 0$ in units of the reciprocal lattice vector $c^*$. The trajectory contained a total of 91 loops.
\begin{verbatim}
	# Loop (z)       BerryPhase(BP)   BP(-/+pi wrap)   BP(unwrap)
	1.00000          -6.28319          -0.00000          -0.00000
	0.98889          0.00712          0.00712          0.00712
	0.97778          0.01377          0.01377          0.01377
	0.96667          0.02049          0.02049          0.02049
	0.95556          0.02738          0.02738          0.02738
	0.94444          -6.24853          0.03466          0.03466
	0.93333          6.32555          0.04236          0.04236
	0.92222          0.05028          0.05028          0.05028
	0.91111          0.05912          0.05912          0.05912
	0.90000          0.06868          0.06868          0.06868
	...              ...              ...              ...
\end{verbatim}

\subsection{Sample work flow}

\begin{figure}
	\includegraphics{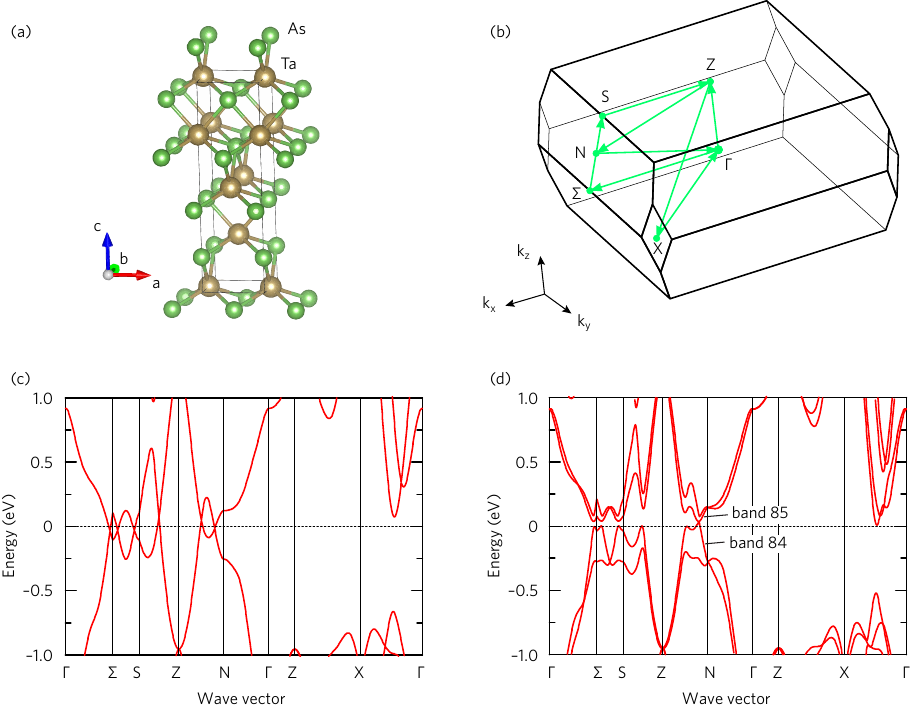}
	\caption{(a) The crystal structure of \ce{TaAs}. (b) The bulk Brillouin zone. (c) Band structure calculated with the PBE functional without SOC. Energies are plotted relative to the Fermi energy. (d) Band structure with SOC. The Berry phase calculations include band 84 only.}
	\label{fgr:Figure-4}
\end{figure}

Here we present a sample calculation for TaAs using \texttt{WIEN2k} with the \citet*{PhysRevLett.77.3865} (PBE) exchange-correlation functional. The steps are as follows:
\begin{itemize}
	\item Generate the input structure file for TaAs using the \texttt{w2web} interface, the \texttt{makestruct} utility or from a cif file. The following lattice parameters for TaAs were used: $a = b = 3.436$~{\AA}; $c = 11.640$~{\AA} and angles $\alpha = \beta = \gamma = 90^\circ$. \ce{TaAs} crystallizes in a body-centered-tetragonal structure with space group $I4_1md$ (No. 109). The unit cell contains 2 non-equivalent atoms with  fractional coordinates Ta $(0, 0, 0.75)$, and As $(0, 0, 0.1677)$ \cite{PhysRevX.5.011029}. The structure of TaAs is shown in Fig.~\ref{fgr:Figure-4}a.

	\item Initialize the calculation with PBE, $3\%$ reduction of muffin-tin radii $R_\text{MT}$, the product $\text{min}(R_\text{MT})K_\text{max} = 7$, and 300 $k$ points in the whole BZ.
	\begin{verbatim}
		init_lapw -b -vxc 13 -red 3 -rkmax 7 -numk 300
	\end{verbatim}
	
	\item Perform a self-consistent field (SCF) calculation with energy convergence of $10^{-4}$~Ry, and charge convergence of $10^{-3}e$.
	\begin{verbatim}
	run_lapw -ec 0.0001 -cc 0.001
	\end{verbatim}
	
	\item Save the calculation to the folder \texttt{noSOC}.
	\begin{verbatim}
	save_lapw -d noSOC
	\end{verbatim}
	
	\item Initialize the SOC using all the default parameters and run a SCF cycle with SOC.
	\begin{verbatim}
	init_so_lapw
	run_lapw -ec 0.0001 -cc 0.001 -so
	\end{verbatim}
	
	\item Calculate the band structure to help locate band crossings for potential Weyl points. Generate a $\Gamma-\Sigma-S-Z-N-\Gamma-Z-X-\Gamma \ k$ path with 600 intermediate points using XCrysDen \cite{KOKALJ1999176}, and save the $k$ point list to a file \texttt{case.klist\_band} (here the case is TaAs, so the file name is \texttt{TaAs.klist\_band}). Afterward, check the \texttt{case.klist\_band} for entries like ``******" that show a problem with format. If this formatting error appears, the number of points on the path needs to be reduced.
		
	\item Calculate the eigenvalues and wave functions for all $k$ points selected on the path.
	\begin{verbatim}
	x lapw1 -band
	x lapwso
	\end{verbatim}
	
	\item Generate files for plotting the band structure, \texttt{case.spaghetti\_ene}.
	\begin{verbatim}
	x spaghetti -so
	\end{verbatim}
	
	\item Plot the band structures without SOC and with SOC (Fig.~\ref{fgr:Figure-5}c,d). For obtaining band structure data without SOC one has to run spaghetti program without \texttt{-so} option (\texttt{x spaghetti}) after the command \texttt{x lapw1 -band} and save the data for plotting. There is a small gap of a few meV in the vicinity of the Weyl points because our path does not intersect the band crossing precisely.
	
	\item To calculate the chirality of the Weyl point, make an input file for the \texttt{WloopPHI.py} code. For TaAs, the input file is given below:
	\begin{verbatim}
	91
	84:84
	&WloopCoordinate
	0.25000 0.00000 1.00000 ; 0.25000 0.00000 0.00000
	0.29000 0.00000 1.00000 ; 0.29000 0.00000 0.00000
	0.30000 0.04000 1.00000 ; 0.30000 0.04000 0.00000
	0.29000 0.08000 1.00000 ; 0.29000 0.08000 0.00000
	0.25000 0.08000 1.00000 ; 0.25000 0.08000 0.00000
	0.24000 0.04000 1.00000 ; 0.24000 0.04000 0.00000
	END
	\end{verbatim}
	The description of input file line-by-line is given below:
	\begin{verbatim}
		Line 1: 91
	\end{verbatim}
		This line gives the number of Wilson loops including the initial point and the final point on the trajectory.		
	\begin{verbatim}
		Line 2: 84:84
	\end{verbatim}
		It specifies the range of bands $n_a:n_b$ in Eq.~\eqref{eq:Berry-flux-numerical}. In \ce{TaAs} (Fig.~\ref{fgr:Figure-4}d) crossings occur between bands 84 and 85, hence only band 84 (lower band of the Weyl point, $n_a=n_b=N$) is considered. The chirality would switch the sign to opposite, if we were to select the upper band instead ($n_a=n_b=N+1$). The phase vanishes if both bands are selected ($n_a=N$, $n_b=N+1$). It is possible to select all underlying bands \texttt{1:84} if we suspect crossings between bands 83 and 84 might enter into the Wilson loop and interfere with the Weyl point in question.
	\begin{verbatim}
		Line 3: &WloopCoordinate
	\end{verbatim}
		This line indicates the beginning of the Wilson loop coordinate list. 
	\begin{verbatim}
		Lines 4-9:
		0.25000 0.00000 1.00000 ; 0.25000 0.00000 0.00000
		...
	\end{verbatim}
		Each line contains the coordinates of a point on the initial Wilson loop and the final Wilson loop of the trajectory. Here, we define a hexagonal Wilson loop (thus six lines) whose starting coordinates $k_{x}$, $k_{y}$, and $k_{z}$ are given first, before the semicolon, and the ending coordinates are given after the semicolon. Coordinates of the $k$ points in the \textit{conventional} (neither primitive, nor Cartesian) BZ can be extracted from \texttt{XCrysDen}. This loop coordinates are chosen in such a way that it encloses the band crossing at coordinate $(k_x=0.28, k_y=0.02, k_z=0.59)$, which we will analyze concerning the chirality. The Wilson loop trajectory enclosing the Weyl point evolves along the $k_{z}$ direction because the Weyl point which we want to characterize is located along the $k$-path Z--N in Fig.~\ref{fgr:Figure-4}c. One can also use more $k$ points to construct the Wilson loop by adding more lines to this section of the input file.
	\begin{verbatim}
		Line 10: END
	\end{verbatim}
		This the end of the input. It should always be written in the input file.
	
	\item Execute \texttt{WloopPHI}
	\begin{verbatim}
		python /path/to/WIEN2k/SRC_BerryPI/BerryPI/WloopPHI.py Wloop.in [-sp] [-orb] [-p]
	\end{verbatim}
	This is the most computationally intensive part. The \texttt{WloopPHI} script automatically executes the following steps for each Wilson loop along the trajectory: (i) generate a series of \texttt{case.klist} with coordinates of $k$ points on the loop interpolated between the initial and final $k$ points and (ii) run  \texttt{BerryPI} to calculate the Berry phase $\phi$ on each loop. The second step includes execution of \texttt{WIEN2k} to generate wave functions for $k$ on the loop and \texttt{Wien2wannier} to calculate the overlap matrix $\mathbb{M}$. The \texttt{WloopPHI} output data are tabulated in the \texttt{PHI.dat} file.
	
	\item Plot the data from \texttt{PHI.dat} and calculate the chirality using Eq.~\eqref{eq:chirality-num}.
	
\end{itemize}

\subsection{Convergence and reliability tests}

The general strategy for finding Weyl points is to first analyze the \textit{non-relativistic} band structure for the band crossings which hold a two-fold degeneracy (excluding spin). These crossings belong to continuous nodal lines that can be mapped within the BZ. Nodal lines of interest are those intersecting the Fermi energy. Weyl points are located in the proximity to these nodal lines. Finding exact positions of Weyl points is done by inspecting band crossings on \textit{relativistic} band structures using off-symmetry $k$ point trajectories. Afterwards, we construct the Wilson loop around the band crossings of interest and find the evolution of the Berry curvature along the Wilson loop trajectory which gives the information on the chirality.

\begin{figure}
	\includegraphics{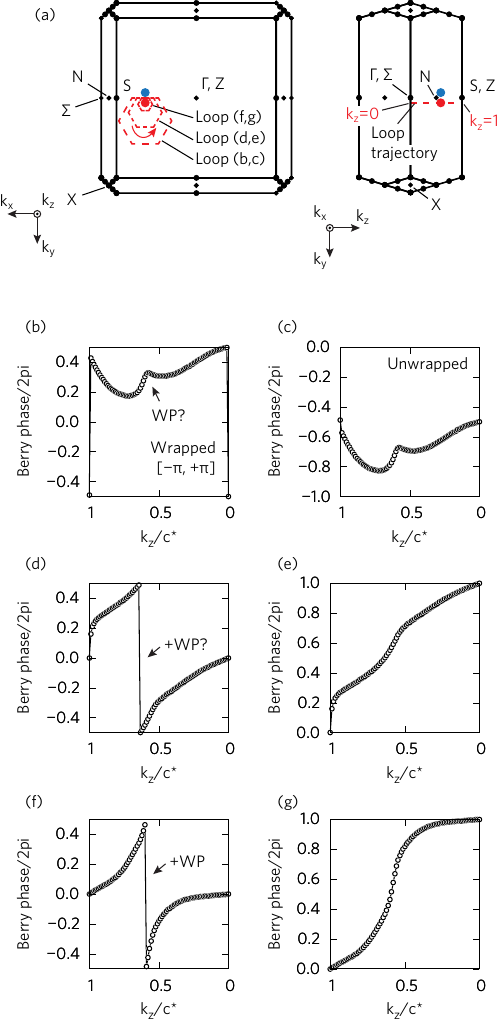}
	\caption{Effect of the Wilson loop size on evaluation of the Weyl node chirality in TaAs. (a) Brillouin zone with loops of various sizes (large, intermediate, and small). (b,d,f) Evolution of the Berry phase (wrapped) for large, intermediate, and small loops, respectively. (c,e,d) The same for the unwrapped Berry phase. The Weyl point identification using the largest loop has failed since other neighbouring Weyl point also contribute to the Berry flux and obstruct the result. Smaller loop sizes are preferable to achieve reliable data.}
	\label{fgr:Figure-2}
\end{figure}

Calculations of the chirality are sensitive to two parameters: the size of the Wilson loop and the number of loops along the trajectory. Figure~\ref{fgr:Figure-2} shows the Wilson loop size effect on the computed Berry phase and the chirality. Here, we considered three cases of the loop size (large, medium, and small) with the same trajectory along $k_{z}$, which encloses one of the Weyl points in TaAs (Fig.~\ref{fgr:Figure-2}a). Figure~\ref{fgr:Figure-2}b shows the effect of a large loop size, where the computed Berry phase exhibits a smooth evolution along the trajectory without any discontinuity, even though the trajectory contains one Weyl point. Due to the large loop size, other Weyl points located nearby induce a non-negligible flux through the loop and interfere with the selected point. Thus, we are unable to characterize the Weyl point and obtain a wrong chirality ($\chi = 0$), as shown in Fig.~\ref{fgr:Figure-2}c. Starting with a medium loop size it becomes possible to characterize the chirality of the Weyl point (Fig.~\ref{fgr:Figure-2}d,e). The smaller is the loop, the easier to identify a discontinuity in the Berry phase and the more accurate is the chirality (Fig.~\ref{fgr:Figure-2}f,g).

\begin{figure}
	\includegraphics{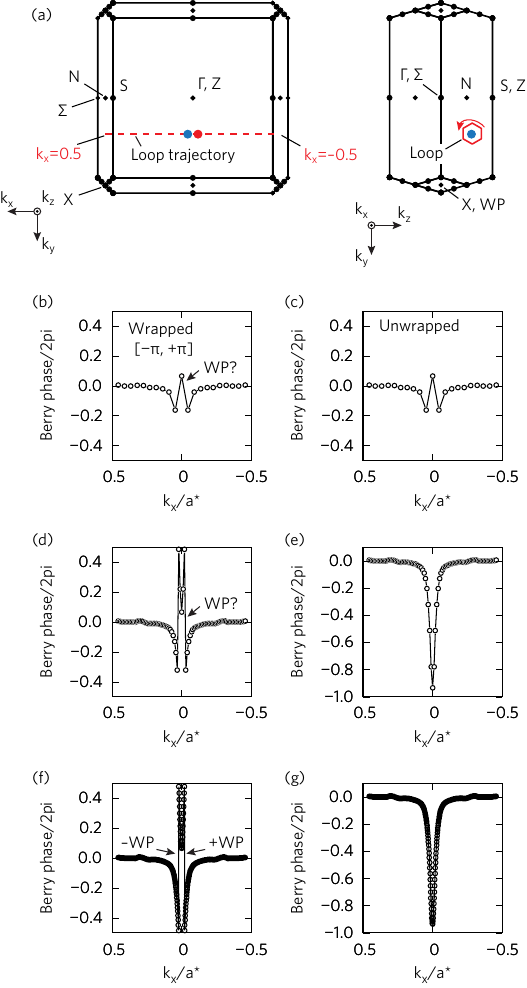}
	\caption{Effect of the number of intermediate Wilson loops along the loop trajectory on evaluation of the Weyl node chirality in TaAs: (a) Brillouin zone with Weyl nodes (only two relevant ones are shown), Wilson loop and its trajectory. (b,c), (d,e), (f,g) Evolution of wrapped and unwrapped Berry phase $\phi(k)$ along the loop trajectory sampled with 21, 91 and 551 equally spaced Wilson loops, respectively.  The data get smoother and more reliable as the number of Wilson loops increases. Only with the highest mesh number we can clearly resolve two $2\pi$ discontinuities on the panel (f), which correspond to two Weyl points of opposite chirality $\chi=-1$ (left) and $\chi=+1$ (right).}
	\label{fgr:Figure-3}
\end{figure}

The effect of the number of intermediate Wilson loops along the trajectory path on characterization of a doublet of Weyl points in TaAs is shown in Fig.~\ref{fgr:Figure-3}. For comparison, we have considered three cases with 21, 91, and 551 Wilson loops. This time we selected a doublet of closely-spaced Weyl points in $k_{x}$ direction (Fig.~\ref{fgr:Figure-3}a) of initially unknown chirality. If we compare the results in Fig.~\ref{fgr:Figure-3}b,d,f, it is clearly shown that when we increase the number of Wilson loops, the evolution of the Berry phase becomes sharper allowing us to resolve two discontinuities, \textit{i.e.}, two Weyl points. Since the unwrapped phase first drops and then raises again (Fig.~\ref{fgr:Figure-5}g), we can conclude that the two Weyl points have different polarity and project no chirality on the 2D (100) surface BZ.

\section{Results}

\subsection{TaAs: method verification}

To validate the method, we analyze the Weyl points of \ce{TaAs}, a well-characterized topological material \cite{PhysRevX.5.011029,Xu613,PhysRevX.5.031013}.  First, we calculate the band structure of \ce{TaAs} without SOC with the PBE functional along high-symmetry directions in the BZ, as shown in Fig.~\ref{fgr:Figure-4}b,c. The band structure plot without SOC (Fig.~\ref{fgr:Figure-4}c) suggests that  multiple band crossings (Dirac nodes) occur close to the Fermi energy along the $\Gamma$--$\Sigma$, $\Sigma$--S, S--Z, and Z--N lines. Our results are in good agreement with previous studies \cite{Xu613,PhysRevX.5.011029}. For the band structure with SOC (Fig.~\ref{fgr:Figure-4}d), we find that all the band crossings near the Fermi energy disappear. Instead, there is a small band gap of around 5~meV along the Z--N line. We can explore the vicinity of this $k$ point to see if it includes any Weyl points.

\begin{figure}
	\includegraphics{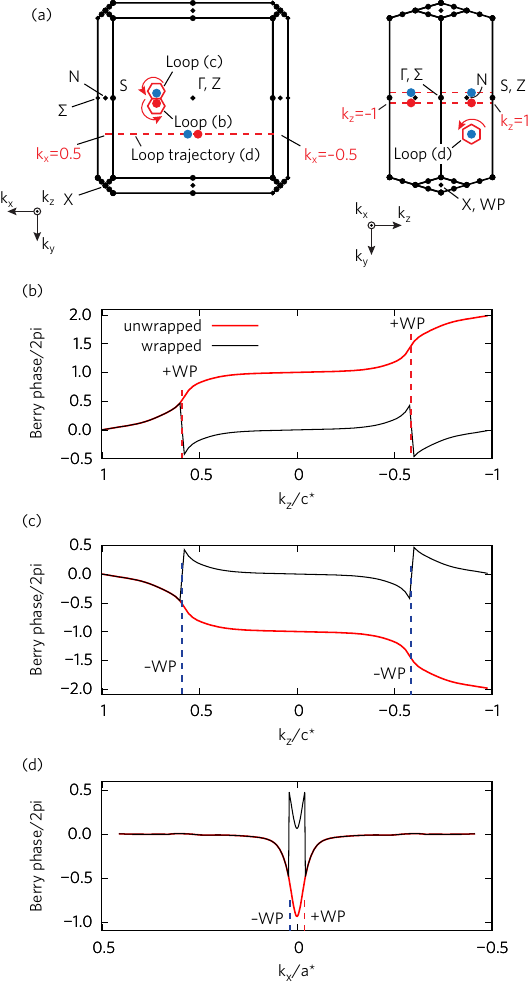}
	\caption{Characterization Weyl points in \ce{TaAs} using the extended Wilson loop method. (a) BZ projections with Weyl points in questions. The loops and their trajectories are shown. Arrows indicate the order in which $k$ points are assembled in the loop. (b,c,d) Evolution of the Berry flux through three different loops along their trajectories. (b) Two positive Weyl points. (c) Two negative Weyl points. (d) A pair of Weyl points of opposite chirality.}
	\label{fgr:Figure-5}
\end{figure}

\begin{figure}
	\includegraphics{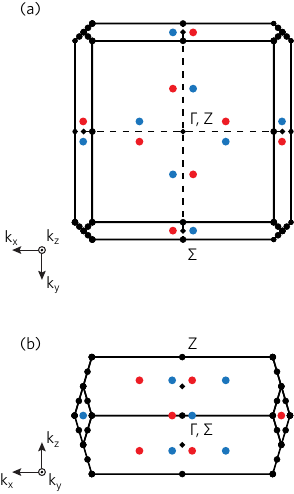}
	\caption{Mapping of Weyl points in the BZ of \ce{TaAs}. (a,b) Top and side view of BZ. There are in total 12 pairs of Weyl points with $+1$ and $-1$ chirality marked as red and blue circles, respectively. }
	\label{fgr:Figure-6}
\end{figure}

To calculate the chirality of the Weyl point formed by the band crossing along the Z--N direction in the  BZ, we construct a Wilson loop which evolves from $k_z=1$ to $k_z = -1$ in units of the conventional reciprocal lattice vector $c^*$ (Fig.~\ref{fgr:Figure-5}a). The result for loop (b) is plotted in Fig.~\ref{fgr:Figure-5}b. It shows two Weyl points of chirality $\chi = +1$ each, which is an odd number (non-trivial), and hence it confirms the existence of Weyl points in  \ce{TaAs}.  Similarly, we apply our methodology to characterize other Weyl points in the BZ using loops (c) and (d) (see Fig.~\ref{fgr:Figure-5}c,d), which include another pair of points with $\chi=-1$ and a pair of points with the opposite chirality. There are a total of 6 Weyl points in the irreducible BZ, which translates to 24 Weyl points in the whole BZ due to the fourfold rotation symmetry (Fig.~\ref{fgr:Figure-6}). This result is in agreement with previous studies \cite{PhysRevX.5.011029,Xu613,PhysRevX.5.031013}.

\subsection{\ce{YRh6Ge4}}

Now that our method is validated, we can examine a  recently proposed topological semimetal \ce{YRh6Ge4} \cite{doi:10.1002/zaac.201300369,PhysRevB.101.035133,PhysRevB.98.045134}. \ce{YRh6Ge4} belongs to the $P\bar{6}m2$ space group with lattice parameters $a = b = 7.067(3)$~{\AA} and $c = 3.862(2)$~{\AA} \cite{PhysRevB.101.035133} and angles $\alpha = \beta = 90^\circ$ and $\gamma = 120^\circ$. The structure optimization was performed with the projected augmented wave method \cite{PhysRevB.50.17953,PhysRevB.59.1758} as implemented in the Vienna \textit{ab initio} simulation package package \cite{PhysRevB.47.558,KRESSE199615,PhysRevB.54.11169}. The PBE version of a generalized gradient approximation was used for the exchange-correlation functional. Pseudopotentials with semicore electrons were selected for Y, Rh, Ge and included 11, 15, 14 valence electrons, respectively. We used a kinetic energy cutoff of 390~eV for the plane-wave basis (25\% higher than recommended) and the $5 \times 5 \times 8$ $k$-point mesh for BZ sampling. Both the cell parameters and internal atomic positions were fully relaxed until the forces on all atoms were smaller than 0.02~eV/{\AA}. The optimized parameters were $a = b = 7.186$~{\AA} and $c = 3.851$~{\AA}. Once the equilibrium structure was obtained, the electronic structure calculation was performed with \texttt{WIEN2k} (the plane wave cutoff parameter of $RK_{\text{max}}=8$ and $5 \times 5 \times 10$ $k$-point mesh for BZ sampling).

\begin{figure}
	\includegraphics{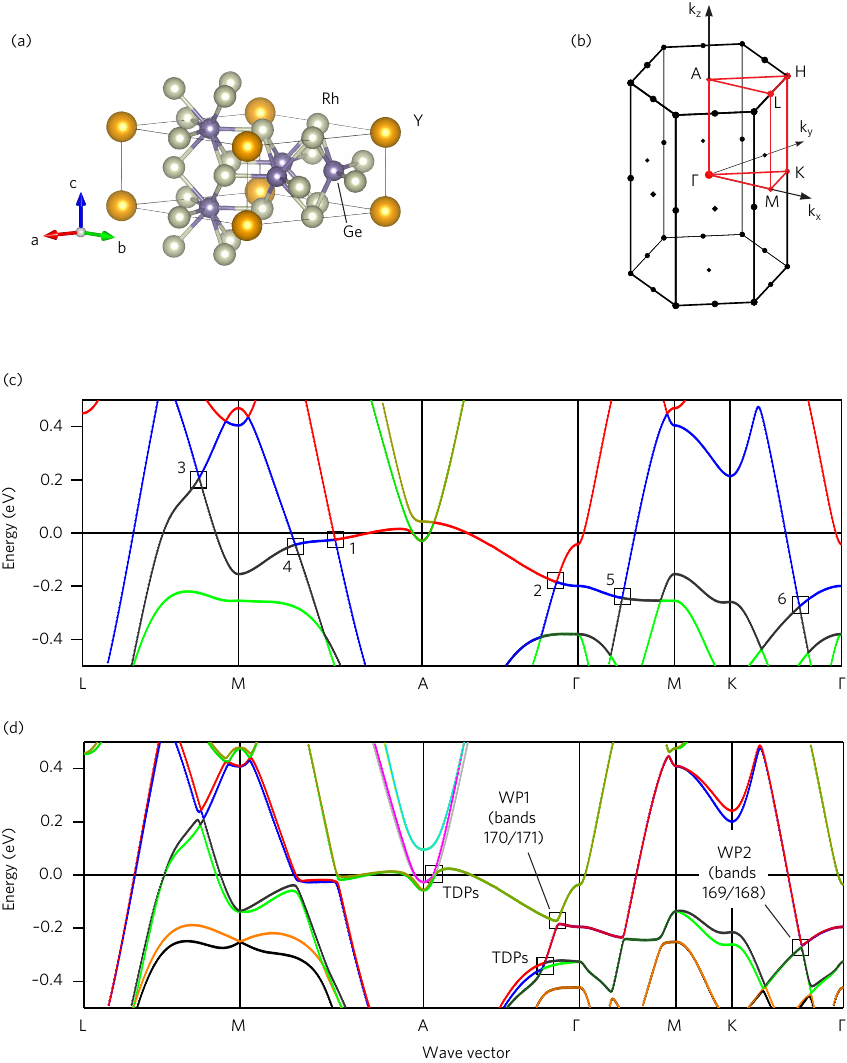}
	\caption{(a) Crystal structure of \ce{YRh6Ge4}. (b) Bulk BZ. (c) Band structure calculated with the PBE functional without SOC. Band crossings with two-fold degeneracies (excluding spin) are labelled $1-6$. They belong to nodal lines that cross the Fermi energy. Energies are plotted relative to the Fermi energy. (d) Band structure with SOC. The proximity to two Weyl points is identified on a path between high-symmetry $k$ points. TDPs refers to a pair of triply degenerate points.}
	\label{fgr:Figure-7}
\end{figure}

\begin{figure}
	\includegraphics{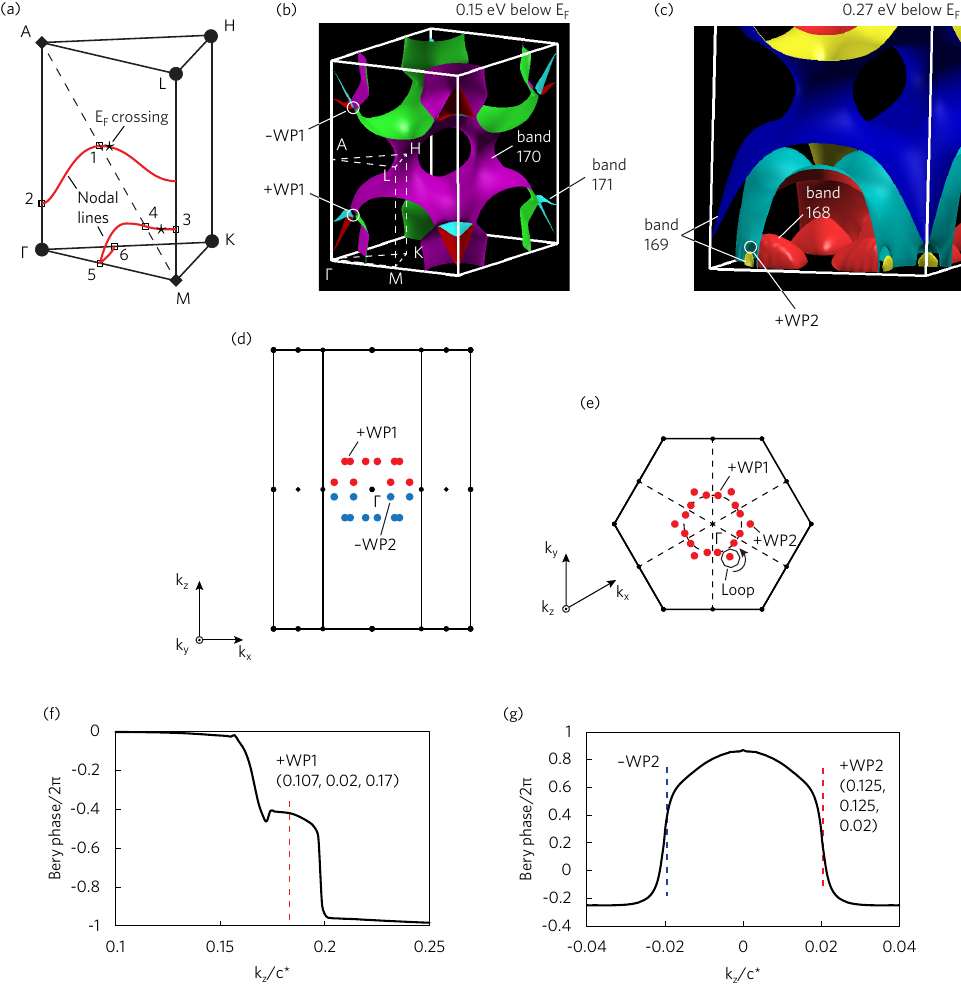}
	\caption{Mapping of Weyl points in the Brillouin zone of \ce{YRh6Ge4}: (a) Irreducible Brillouin zone with two nodal lines (without SOC) that intercept the Fermi energy at points marked with asterisks. Points $1-6$ on the nodal lines correspond to nodal crossings labelled in Fig.~\ref{fgr:Figure-7}c. (b,c) Touching points on relativistic energy isosurfaces are identified as Weyl points (WP1 and WP2). (e,d) The top and side view of the Brillouin zone. Weyl points are shown with red and blue solid circles. (f,g) Evolution of the Berry flux through the Wilson loops along its trajectory allows to identify a $k_z$ coordinate, the sign and chirality of Weyl points.}
	\label{fgr:Figure-8}
\end{figure}

Figure~\ref{fgr:Figure-7}a,b shows the crystal structure and bulk BZ  of \ce{YRh6Ge4}, respectively. The calculated band structure of \ce{YRh6Ge4} without SOC is presented in Fig.~\ref{fgr:Figure-7}c. We can see multiple band crossings $1-6$ associated with two nodal lines shown in Fig.~\ref{fgr:Figure-8}a (not to be confused with nodal lines in the band structure \textit{with} SOC discussed later). These are the only nodal lines that cross the Fermi energy and, thus, both are of experimental importance.

The search for Weyl points begins with inspecting a relativistic band structure along the same $k$ path (Fig.~\ref{fgr:Figure-7}d). Nodal points are now gapped due to the lack of inversion symmetry. The smallest gap emerges at the band crossing along the $\Gamma$--K direction in the BZ (labelled as WP2 on Fig.~\ref{fgr:Figure-7}d). The location of WP2 in the BZ is illustrated in Fig.~\ref{fgr:Figure-8}c as a touching point between energy isosurfaces (0.27~eV below the Fermi energy) of two bands 168 and 169. To evaluate the chirality of the WP2 Weyl points, we construct the Wilson loop with a trajectory from  $k_z = -0.04$ to $k_z = +0.04$ plane (Fig.~\ref{fgr:Figure-8}e,d). We choose the Wilson loop trajectory in such a way that it includes the pair of Weyl points WP2 along the $\Gamma$--K direction. The Berry phase evolution along the Wilson loop trajectory is plotted in Fig.~\ref{fgr:Figure-8}g. One can identify two places of sharp increase and decrease in $\phi(k_z)$. The plot is similar to Fig.~\ref{fgr:Figure-5}d. Further analysis of the unwrapped Berry phase yields the chirality of the WP2 Weyl points $\chi = -1$ and $\chi = +1$ labelled on Fig.~\ref{fgr:Figure-8}e,d.

The second nodal line in Fig.~\ref{fgr:Figure-8}a also contributes another potential Weyl point located not far from $\Gamma$--A direction (labelled as WP1 on Fig.~\ref{fgr:Figure-7}d). Its location in the BZ is illustrated in Fig.~\ref{fgr:Figure-8}b as a touching point between energy isosurfaces (0.15~eV below the Fermi energy) of two bands 170 and 171. Similarly to WP2, we construct a Wilson loop that encloses WP1 and translate the loop along $k_z$ direction. Evolution of the berry phase presented in Fig.~\ref{fgr:Figure-8}f shows a positive chirality for the point located at $k_z \approx 0.17$. The mirror point at $k_z \approx -0.17$ has an opposite chirality. Figure~\ref{fgr:Figure-8}e,d maps sets of WP1 and WP2 points with their chirality color-coded within the conventional hexagonal BZ. The chirality of Weyl points cancels out if projected on the $(0001)$ plane, but it is not the case for the $(10\bar{1}0)$ surface.

\begin{figure}
	\includegraphics{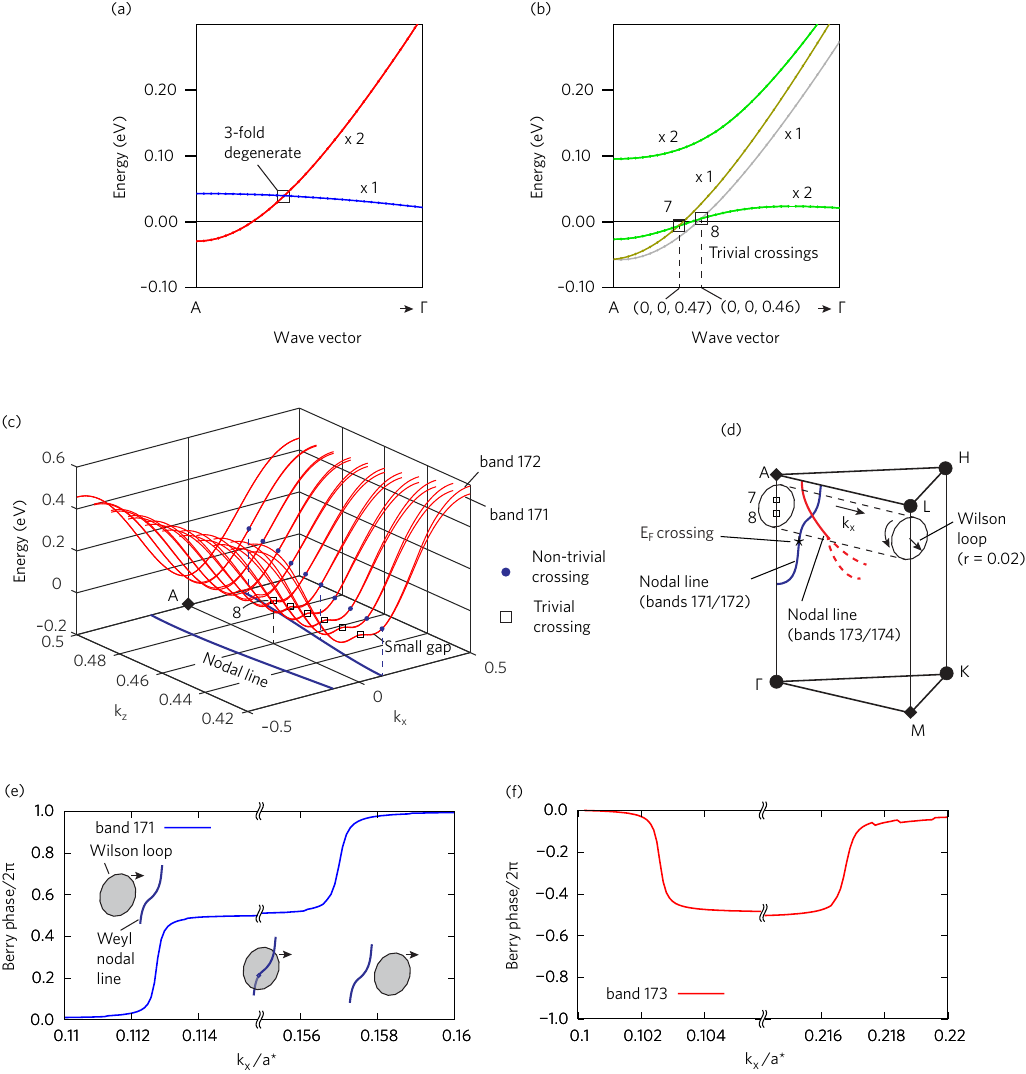}
	\caption{Triply degenerate points and nodal lines in \ce{YRh6Ge4}: (a,b) Band structure without and with SOC (colours reflect different irreducible representations). Degeneracy of bands is indicated (excluding spin on panel (a)). Crossings labeled 7 and 8 are trivial since bands lie in the mirror plane. (c) Band structure with SOC in direction perpendicular to $A-\Gamma$ shows non-trivial band crossings that form continuous nodal lines. (d) Wilson loop trajectory. (e,f) Berry phase along the Wilson loop trajectories reveals chirality of nodal lines.}
	\label{fgr:Figure-9}
\end{figure}

\citet{PhysRevB.98.045134} emphasized the presence of triply degenerate nodal points (TDPs) in the band structure of \ce{YRh6Ge4} (also marked in Fig.~\ref{fgr:Figure-7}d). Of particular interest are the TDP crossings 7 and 8 near the A point, which are very close to the Fermi energy (see a close-up view of the band structure in Fig.~\ref{fgr:Figure-9}a,b). Because of $D_{3h}$ point group symmetry, these crossings occur at the high symmetry axis $\Gamma-A$ which lie in the mirror plane $A-\Gamma-H$ of BZ in Fig.~\ref{fgr:Figure-7}b \cite{barman_symmetry-driven_2020}. Due to $D_{3h}$ point group symmetry the $C_3$ eigenvalues of the bands are degenerate which results in the trivial pair of TDPs \cite{zhu2016triple}. We inspected the chirality of TDPs 7 and 8 (Fig.~\ref{fgr:Figure-9}b) by constructing a Wilson loop around the crossings and found $\chi(k_z) = 0$ for each crossing band individually thereby confirming the trivial topology of those TDPs.

While inspecting the band dispersion in the proximity of TDPs, we discovered another set of non-trivial band crossings shown in Fig.~\ref{fgr:Figure-9}c. With SOC these band crossings form continuous nodal lines. This happens because the crossings lie on the mirror plane $A-\Gamma-M$ (Fig.~\ref{fgr:Figure-9}d) which protects and make them robust against SOC \cite{jin_two-dimensional_2020}. Here we observe a nodal line that passes through the Fermi energy (Fig.~\ref{fgr:Figure-9}d). A Berry phase of $\pm\pi$ is anticipated for any loop that interlocks with a nodal line \cite{Fang_PRB_92_2015,Bian_NC_7_2016}. Even though it is possible to construct a Wilson loop that interlocks the nodal line and determine the Berry phase in a single calculation, it will not be possible to discern between $+\pi$ and $-\pi$ chirality due to the $\pi$-wrapping of the phase. To observe the evolution of the phase, we choose a strategy similar to the characterization of Weyl points. We calculate the evolution of the Berry phase on the Wilson loop as a function of its trajectory selected such that its path will intersect the nodal line (Fig.~\ref{fgr:Figure-9}d). The Wilson loop analysis (Fig.~\ref{fgr:Figure-9}e,f) confirms the non-trivial chirality of nodal lines. The Berry phase on the loop changes abruptly by the magnitude of $\pi$ once the nodal line enters the surface of the loop (see inset in Fig.~\ref{fgr:Figure-9}e). The second step in $\chi(k_x)$ occurs as the nodal line exits the Wilson loop surface. This two-step behavior is different from Weyl points. It is possible that the Berry phase profile in Fig.~\ref{fgr:Figure-8}f is also linked to nodal lines.

It should be noted that the position of the band crossings in \ce{YRh6Ge4} is sensitive to lattice parameters. In our calculation PBE self-consistent optimized lattice parameters were used, which are within 2\% of their experimental counterparts. If experimental lattice parameters were used instead (with only atomic positions relaxed at the PBE level), the nodal crossing 2 in Fig.~\ref{fgr:Figure-7}c would move to the $\Gamma$--M section of the $k$ path (close to $\Gamma$ point) without significant changes in its position relative to the Fermi energy.

\section{Conclusion}

We described the methodology of \texttt{WloopPHI}, a code designed to calculate the chirality of band crossing points for the characterization of Weyl points using an extended Wilson loop method. The code \texttt{WloopPHI} has been implemented in WIEN2k (an all-electron density functional package). Our method is verified using \ce{TaAs}---a well-known Weyl semimetal---and the results show good agreement with previous studies. Further, we have applied the approach to the newly studied material \ce{YRh6Ge4}. Its band structure without the spin-orbit coupling exhibits two nodal lines that cross the Fermi energy. These nodal lines transform into two sets of Weyl points (located $\sim 0.2$~eV below the Fermi energy) when the spin-orbit coupling is included. The chirality (or monopole charge) of Weyl points is linked to a change in the Berry phase accumulated on the Wilson loop $(\pm2\pi)$ as the point in question passes through the surface of the loop. We mapped chirality and location in the Brillouin zone for Weyl points in the proximity of the Fermi energy. In addition, nodal lines were observed in the relativistic band structure. These nodal lines are robust against SOC because of horizontal mirror symmetry. Of particular interest is one of them that crosses the Fermi energy. A distinct feature of the nodal line (in contrast to the Weyl points) is the presence of two steps of the magnitude $\pm\pi$ in the Berry phase on the Wilson loop associated with the nodal line entering the surface of the loop and then exiting it. The obtained results show that it is feasible to calculate the chirality of band crossings with the \texttt{WloopPHI} code.

\begin{acknowledgments}
Authors are indebted to Prof.~Zhiqiang Mao (Pennsylvania State University) for drawing out attention to the  \ce{YRh6Ge4} material. Funding was provided by the Natural Sciences and Engineering Research Council of Canada (NSERC) under the Discovery Grant Program RGPIN-2020-04788. Calculations were performed using a Compute Canada infrastructure supported by the Canada Foundation for Innovation under the John R. Evans Leaders Fund program.
\end{acknowledgments}


%


\clearpage

\end{document}